\documentstyle [12pt]{article}
\newcommand{\be}{\begin{equation}}
\newcommand{\ee}{\end{equation}}
\newcommand{\bea}{\begin{eqnarray}}
\newcommand{\eea}{\end{eqnarray}}
\newcommand{\ba}{\begin{array}}
\newcommand{\ea}{\end{array}}

\begin{document}
\baselineskip = 19 pt
 \thispagestyle{empty}
 \title{
\vspace*{-2.5cm}
\begin{flushright}
\begin{tabular}{c c c c}
\vspace{-0.3cm}
& {\normalsize MPI-Ph/93-58}\\
\vspace{-0.3cm}
& {\normalsize SSCL-Preprint-497}\\
\vspace{-0.3cm}
& {\normalsize August 1993}
\end{tabular}
\end{flushright}
\vspace{1.5cm}
Infrared  Fixed Point Solution for the Top Quark Mass
and Unification of Couplings in the MSSM
\\
 ~\\}
 \author{
 W. A. Bardeen$^{a}$, $\;$
 M. Carena$^{b}$,\\
{}~\\
S. Pokorski$^{b}$\thanks{On leave from the Institute of Theoretical
Physics,
 Warsaw University}
        $\;$ and  C. E. M. Wagner$^b$\\
 ~\\
{}~\\
$^{(a)}$Theoretical Physics, SSC Laboratory,\\
2550 Beckleymeade Ave., Dallas, Texas 75237-3946, U.S.A.\\
{}~\\
 ~\\
$^{(b)}$Max-Planck-Institut f\"{u}r Physik, Werner-Heisenberg-Institut\\
F\"{o}hringer Ring 6, D-80805 M\"{u}nchen , Germany.\\
{}~\\
 }
\date{
\begin{abstract}
We analyze the implications of the infrared quasi fixed point
solution for the top quark mass in the Minimal Supersymmetric
Standard Model.  This solution could explain in a natural way
the relatively large value of the top quark mass and, if
confirmed experimentally, may   be  suggestive of the
onset of nonperturbative physics at very
high energy scales.
In the framework of grand unification, the expected bottom quark -
tau lepton Yukawa coupling unification is very sensitive to the fixed
point structure of the top quark mass. For the presently allowed values
of the electroweak parameters and the bottom quark mass, the Yukawa
coupling unification implies that  the top quark mass must be
within ten percent of its fixed point values.
\end{abstract}}
\maketitle
\newpage
In the present expectation of  a very heavy top quark,
we witness a revival of
interest in the infrared quasi fixed point predictions for
the top quark mass.  The idea is that the top
quark mass may be completely determined by the low
energy fixed point structure of the Renormalization
Group (RG)
equations independent of the precise symmetry conditions
at a   high mass scale.
For several reasons, this idea is very appealing.
The infrared stable fixed point structure  of the Standard Model was
first  analysed by Pendleton  and Ross \cite{PR}. Their analysis
focussed on an exact fixed point relationship between the top
quark Yukawa coupling and the QCD coupling which requires
a smooth running  of the couplings to infinite energy or implies
analytic  relations between  the couplings. This approach is closely
related
to the coupling reduction methods advocated by Zimmermann et al.
\cite{Z}.
At difference with the Pendleton-Ross fixed point, the quasi
fixed point structure of C. T. Hill \cite{IR} results from a strong
 focussing of the running  of the top quark Yukawa at low energies for
 sufficiently strong Yukawa couplings at a finite high energy
scale. This behavior reflects the existence of Landau poles
 or nonanalytic relations between the couplings in the RG evolution
somewhat above the high energy scale, where new physics is expected to
control  the actual values of the couplings. A wide range of couplings
at the high energy scale
will fall within the domain of attraction  of the
 quasi fixed point and evolve to a sharply defined  fixed point value
for the top quark mass. The fixed point structure of the Standard Model
provides a natural explanation of larger values of the top quark mass.

In the Minimal Supersymmetric Standard Model (MSSM),
a similar quasi fixed point behavior of the   renormalization group
 solutions is present\cite{Dyn}. In fact, for a given value of
the strong gauge coupling  $\alpha_3(M_Z)$,
one predicts
\begin{equation}
M_t \simeq A \sin\beta
\end{equation}
where $M_t$ is the physical top quark mass, the angle $\beta$ is defined
so that $\tan\beta$ is the ratio  of the two Higgs vacuum expectation
values and
$A \simeq 190 - 210 $ GeV for
$\alpha_3(M_Z) = 0.11 - 0.13$ \cite {BABE} - \cite{CPW}.
It is worth stressing that the above  prediction is obtained for a
range of high energy values of $Y_t = h_t^2/4\pi$,  $h_t $ being the
top quark Yukawa coupling, such that $Y_t$ can reach its
perturbativity limit $Y_t \simeq
{\cal{O}}(1)$ at some scale $M_X = 10^{14} - 10^{19}$ GeV
(see Fig. 1). One can therefore
envision the following scenarios, which would lead to the
infrared quasi fixed point prediction for $M_t$:\\
i) The onset of non-perturbative physics at scales below or
of the order   $M_{GUT} = {\cal{O}}(10^{16}$ GeV),  at which
 the  unification of gauge couplings may take place.
This is what happens, for example,
in the supersymmetric extension of the so called top condensate
models \cite{Dyn}.  In this case, the perturbative analysis of
grand unified scenarios would be invalidated and one should
make an analysis of nonperturbative effects before
studying  the precise unification conditions. \\
ii) Perturbative grand unification, with a value of
$0.1 \leq Y_t \leq 1$ at the grand unification scale, followed
by the onset of nonperturbative physics for scales of the
order of $M_{GUT}$.\\
iii) Perturbative theory up to scales of the order of
$M_{PL}$, but with large $Y_t$, close to its
perturbativity limit, and with the possibility of
 new physics at
scales $\mu \geq M_{GUT}$.\\
In summary,
an infrared quasi fixed point value for $M_t$, if confirmed
experimentally, would either
be strongly suggestive of non-perturbative
physics at very high energy scales,
or would call for understanding the relatively large
values, of
order one, of  the top quark Yukawa coupling
at the high energy scales within the perturbative
scenario.

Going further in the aim of computing the values of the
other fermion masses measured experimentally, it is
natural to think about the
possibility of viewing them in the framework
of grand unified theories. In this respect, bottom-tau
Yukawa coupling unification appears naturally  in
many grand unified schemes \cite{Ramond}-\cite{DHR}.
Most interesting is the
fact that to achieve bottom-tau Yukawa unification,
$h_b(M_{GUT}) = h_{\tau}(M_{GUT})$, large values
of the top quark Yukawa coupling are necessary in order
to compensate for the effects of the strong interaction
renormalization. Indeed, most recently, it has been
observed that one is driven close to the
infrared quasi fixed point by the bottom - tau Yukawa
unification requirement \cite{BABE}, \cite{CPW}.
This property is strongly  dependent on the exact values of
the strong gauge coupling as well as on the physical bottom quark mass
 being in the range of experimentally allowed values
$M_b = 4.9 \pm 0.3$ GeV \cite{Partd}.
If, for example, the physical
bottom mass value were
$M_b \leq 3$ GeV, perturbative
unification of
bottom  and tau Yukawa couplings would not be possible for
$\alpha_3(M_Z) = 0.11 - 0.13$, since even a top quark Yukawa
coupling
 at the edge of the validity of the perturbative
expansion  would not be strong enough to contravene the strong gauge
coupling renormalization of the bottom mass.  In fact, for a given
bottom mass, one can define an upper bound on the strong gauge coupling
for which bottom-tau Yukawa unification becomes possible \cite{CPW}.
Surprisingly
enough, for experimentally allowed values of $M_b$, this upper
bound on $\alpha_3(M_Z)$
lies within the range predicted from LEP measurements.
In addition, the experimental upper bound on $M_b$ is also quite
important to account for  the infrared fixed point behaviour of the
 top quark mass.  For larger values of the bottom quark mass,
$M_b \geq 6$ GeV, there would be no necessity of a strong renormalization
effect from the top quark Yukawa coupling.

The above is quite intriguing, since it implies that just
for the experimentally acceptable values of $M_b$ and $\alpha_3(M_Z)$,
bottom-tau Yukawa unification yields large values -of order one- of the
 top Yukawa coupling at the grand unification scale,
 which, however, remain  in the range of validity of perturbation theory.
This is a highly nontrivial property of
 the Minimal Supersymmetric Standard Model and, as we explained above,
it implies that the top quark mass
is strongly focussed to its infrared quasi fixed point value.  Thus, in
this case, the observed top quark mass will be insensitive  to the
actual value of the top quark Yukawa coupling at high energies. It is
precisely  the bottom-tau  Yukawa unification condition which would
be  uniquely sensitive  to the high energy Yukawa coupling of the
top quark.

In the following, we extend and quantify the previous discussion. In
particular, we address the issue of the
proximity to the infrared quasi fixed point prediction for
$M_t$ when requiring $h_b(M_{GUT}) = h_{\tau}(M_{GUT})$.
In Figs. 2.a - 2.d, we consider the two loop renormalization
group
running of the bottom quark and tau lepton Yukawa couplings,
$Y_b = h_b^2/4\pi$ and $Y_{\tau} = h_{\tau}^2/4\pi$, respectively,
for two representative values of $\alpha_3(M_Z)$ = 0.11,
0.125, and the physical
bottom quark mass $M_b = 4.9, 5.2$ GeV, to analyze under
which conditions the unification of these two Yukawa couplings
is possible.
The larger (smaller)
the value of $\alpha_3(M_Z)$ ($M_b$) the larger is
the value of $h_t$ necessary to  achieve unification of $h_b$ and
$h_{\tau}$ in the range, say $10^{15} - 10^{19}$ GeV.
As we discussed above, there is an obvious reason for this: Large
$h_t$ is needed to partially cancel the strong interaction
renormalization of $M_b$, $h_t$ being necessarily strong at scales of
order $M_{GUT}$ for $M_b = 4.9 \pm 0.3$ GeV and large values of
$\alpha_3(M_Z) \geq 0.115$.
At the same
time, for large values of the top quark Yukawa coupling at the
grand unification scale,
$h_t(M_{GUT})$, for which the top mass is close to its quasi
infrared fixed point value, the $h_b - h_{\tau}$ unification
scale becomes extremely sensitive to the actual top quark
mass value. Changing
$M_t$ by 1 - 3 GeV (for fixed $\tan\beta$) can change the
bottom - tau unification scale by several orders of magnitude.
This
implies, as shown in Fig. 2, that for $\alpha_3(M_Z) = 0.125$,
unification of $h_b$ and $h_{\tau}$ in the range $M_{GUT} =
10^{15} - 10^{19}$ GeV only holds for very restrictive values
of the running top quark mass, $m_t$.\footnote{Once the QCD
corrections
are included, the running top quark mass used above is about
 6$\%$ smaller than the physical pole mass
\cite{CPW}.}
For instance, a   $\tan\beta = 3.5$,
a physical bottom mass $M_b = 4.9$ GeV and $\alpha_3(M_Z) = 0.125$
implies
$m_t \simeq 190 \pm 3$ GeV. Similarly, values of
$\alpha_3(M_Z) = 0.12 \; (0.115)$ imply
 $m_t \simeq 186 \pm 3$ GeV
($ 182 \pm 3$ GeV). Moreover, as can be clearly seen
in Fig. 2, for these values of
$\alpha_3(M_Z)$ the predicted top quark mass is quite
insensitive to
the inclusion of small threshold corrections, of the order of
10$\%$, to the bottom-tau
Yukawa coupling  unification condition.
On the contrary, for smaller values of
$\alpha_3(M_Z) \leq 0.110$, unification of $h_b$ and $h_{\tau}$
may, in principle,
 be achieved for a wide range of smaller values of $m_t$.
For example, for $\alpha_3(M_Z) = 0.105$,
$\tan\beta = 3.5$ and considering $M_b = 5.2$
GeV, bottom - tau Yukawa unification is possible for values
of $m_t \simeq 125 - 160$ GeV. However, as we shall show below,
for a top quark mass $M_t > 110$ GeV,
these relatively low values of the strong gauge coupling,
$\alpha_3(M_Z) \leq 0.110$, can not be self consistently
achieved
if unification of gauge couplings is also required for the MSSM.
Similar features are obtained for the whole low and moderate
$\tan \beta$ regime.

Next, we study in more detail how close are we driven to the
infrared fixed point prediction for $M_t$ by the requirement of
$h_b$ - $h_{\tau}$ unification, as a function of the strong gauge
coupling and for several values of the bottom quark mass and
$\tan \beta$.
The $h_b$ - $h_{\tau}$ unification scale is defined by the
running of the gauge couplings. The results are shown
in Fig. 3.a - 3.c.
We see that, for values of the strong gauge
coupling
$\alpha_3(M_Z) \geq 0.115$, and for $M_b \leq 5.2$ GeV,
the top quark mass is within
a ten percent of its infrared quasi fixed point value.
If, instead, the strong gauge coupling were
$\alpha_3(M_Z) < 0.110$,
the top quark mass could be far away from its infrared quasi
fixed point value.  Concerning possible threshold corrections
to the unification of both Yukawa couplings, it is interesting
to remark that a relaxation in the unification condition of
order 10$\%$,
 $h_b(M_{GUT})/h_{\tau}(M_{GUT}) = 0.9$, for a
bottom mass $M_b = 4.9$ GeV gives approximately the same
behaviour as if we consider exact
$h_b-h_{\tau}$ unification but with $M_b = 5.2$ GeV.

It is important to observe that for smaller values of $M_b$ and
larger values of $\alpha_3(M_Z)$ the top quark Yukawa coupling
required for $h_b$ - $h_{\tau}$ unification
may become too large. For a consistent perturbative
treatment of the theory we require $Y_t(M_{GUT}) = h_t^2(M_{GUT})/4\pi
\leq 1$, which implies that the two loop
contribution to the renormalization group evolution of $h_t$
is less than  a 30 $\%$ of the one loop one. As a matter of
fact, observe that  in Fig. 3.
the curves for $M_b = 4.7$ GeV and
$M_b = 4.9$ GeV do not continue up to $\alpha_3(M_Z) = 0.13$,
since the top quark Yukawa coupling at $M_{GUT}$ would then
become  too large
to be consistent with a perturbative analysis \cite{CPW}.

Considering the constraints coming from
the gauge coupling unification \cite{early} - \cite{RR},
predictions  for $\sin^2\theta_W(M_Z)$ are derived
as a function  of the strong gauge coupling $\alpha_3(M_Z)$.
Indeed, the unification
condition implies the following numerical correlation
\begin{equation}
\sin^2\theta_W(M_Z) = 0.2324 -
0.25 \times (\alpha_3(M_Z) - 0.123)
\pm 0.0025 ,
\end{equation}
where the central value corresponds to an effective supersymmetric
threshold scale \cite{LP}, \cite{CPW}
$T_{SUSY} = M_Z$ and
the error $\pm 0.0025$ is the estimated uncertainty in
the prediction arising from possible
supersymmetric threshold corrections
(corresponding to varying $T_{SUSY}$ from
15 GeV to 1 TeV )
and including possible effects from threshold corrections
at the unification scale as well as from
higher dimensional operators.
On the other hand, $\sin^2\theta_W(M_Z)$ is given by the
electroweak parameters $G_F, M_Z, \alpha_{em}$ as a function
of $M_t$ (at the one - loop level) by the formula \cite{LP}:
\begin{equation}
\sin^2\theta_W(M_Z) = 0.2324 - 10^{-7} \times GeV^{-2}
\times \left( M_t^2 -
(138 GeV)^2 \right) \pm 0.0003
\end{equation}
Therefore, the predictions from
the gauge coupling unification agree with experimental data
provided
\begin{equation}
M_t^2 = (138 GeV)^2
+ 10^7 \times GeV^2 \times 0.25 \times
\left(\alpha_3(M_Z)
-0.123 \pm 0.01 \right)
\label{eq:correl}
\end{equation}
The above     $M_t -\alpha_3(M_Z)$ correlation defines a band, whose
upper bound is
$\alpha_3^u (M_Z) \geq 0.13$. Thus, although this upper bound
does not appear explicitly in Fig. 3.,
Eq (\ref{eq:correl}) implies that the
region in $\alpha_3(M_Z)$ to the right of the dash-long dashed curve
is the allowed one.
 Moreover, in Fig. 3 we observe
the intersection of that allowed region
with the $M_t - \alpha_3(M_Z)$
curves which follow from the $h_b - h_{\tau}$ unification
condition. In fact,
for $M_b$ in the range $(4.9 \pm 0.3)$ GeV, the
$h_b - h_{\tau}$ unification and the gauge coupling unification
condition together with the quadratic dependence on $M_t$ of
$\sin^2\theta_W(M_Z)$
(both within the discussed uncertainties) are compatible
with each other only within a restricted range of
$\alpha_3(M_Z)$ and $M_t$, and,
moreover,  push $M_t$ very close
 to its quasi infrared fixed point values.
  Due to the correlation
between $\sin^2\theta_W(M_Z)$ and $M_t$, the effect becomes
more dramatic for larger values of $\tan\beta$, for which
$\sin\beta \simeq 1$, than for values of $\tan\beta$ close to
one. In general,
for $\tan\beta \geq 1$, the strong gauge coupling takes values
$\alpha_3(M_Z) \geq 0.112$ and $M_t$ is at most a 10$\%$ lower
than its infrared quasi fixed point prediction. Indeed, as may
be observed from Fig. 3, the above
lower bound on $\alpha_3(M_Z)$
may only be reached for $\tan\beta \simeq 1$. For
larger values of $\tan\beta$, the lower bound on $\alpha_3(M_Z)$
increases together with the top quark mass, which has then a  stronger
convergence
to its infrared fixed point behavior.

It is important to remark that the above  study is performed in
the region of moderate values of $\tan\beta \leq 30$. In
this region, the top Yukawa coupling at the grand unification
scale depends only weakly on the exact supersymmetric spectrum
but strongly on the effective supersymmetric
threshold scale $T_{SUSY}$ \cite{CPW}
which determines the low energy values of the strong gauge
coupling coming from the requirement of gauge coupling
unification \cite{LP}.
 The present figures were obtained
for $T_{SUSY} = M_Z$, taking the squark masses to be equal to
the $Z$ boson mass. If, while keeping $T_{SUSY}$ fixed,
the squark masses are increased, in the range $M_Z - 1$ TeV,
the top quark mass may
increase in a few GeV but without changing
the physical picture \cite{CPW}.

For larger values of $\tan\beta$,
for which the bottom Yukawa coupling becomes
sufficiently strong to be able to partially cancel    the strong gauge
coupling renormalization effects in its own running, the
$h_b$ - $h_{\tau}$ unification condition no longer requires the
existence of a large top Yukawa coupling \cite{CPW}.
Therefore, the predictions for $M_t$ are no longer strongly constrained
to be close  to its infrared quasi fixed point values. As
shown in Fig. 3.d, already for values of $M_b$ of the order of
its experimental upper bound, large deviations from the infrared
quasi fixed point predictions for $M_t$ are observed, for a wide
range of experimentally allowed values for $\alpha_3(M_Z)$.
In addition, for very large values of $\tan\beta$,
large corrections to the running bottom quark
mass may be present \cite{Ralf},\cite{Hall} ,
and additional symmetries may be
required to cancel them in a natural way. Such symmetries
 may, however, be
 in conflict with the
radiative breaking of $SU(2)_L
\times U(1)_Y$ \cite{OP}.  The loss of
stability for the top quark mass values in the large $\tan \beta$
regime is also reflected in a
 stronger sensitivity to the exact supersymmetric
 spectrum.
Furthermore, in   Fig. 3.d. it is to observe that for $\tan \beta =$ 50
the condition of unification of the three Yukawa couplings,
$h_t(M_{GUT}) =  h_b(M_{GUT}) =  h_{\tau}(M_{GUT}) $ (dot-long dashed
line)
determines a large value  of the top quark mass
which, however
differs from the infrared fixed point behavior. In this case, the
value of the Yukawa couplings at the unification scale
is of the order of the unification  gauge coupling.
In addition, for lower values of the bottom quark mass the predictions
for $M_t$ are outside the range allowed by the
gauge coupling unification condition.
To achieve
unification of the three Yukawa couplings inducing as well values
of the top quark mass close to its infrared fixed
 point, a larger value
of $\tan \beta$ is required.

Finally, concerning the  present experimental limits on the top
quark mass it follows that,  for a Higgs mass
$m_h \simeq 100$ GeV,
 the direct experimental
determination of the Weinberg angle from LEP yields
$M_t \simeq
150 \pm 40$ GeV at the $95 \%$ confidence level \cite{Alta}.
The same result would be obtained in the Minimal
Supersymmetric Standard Model, if all supersymmetric particle masses
 $M_{SUSY} \gg M_Z$. However, as has been recently pointed out,
this result could be modified in case there were light supersymmetric
particles, with masses close to their present experimental
bound \cite{Sola},\cite{ABC}.
In the heavy
MSSM  scenario, with no light supersymmetric particles, the
rather large prediction for the top quark mass coming from the
most recent experimental measurements allows
the accomodation of the
above discussed scenario for a reasonable range of values of
$\tan\beta$ at the two sigma level.
 The light minimal supersymmetric
scenario may  even improve the agreement,  depending on which
supersymmetric particles  become light.

In conclusion, in the
Minimal Supersymmetric Standard Model, for any given value
of the strong gauge coupling $\alpha_3(M_Z)$, the infrared quasi
fixed point solution naturally provides
predictions of a large top quark mass as a function of $\tan\beta$.
We have shown that such  values of the top quark mass may  also be
naturally obtained within the context of grand unified
scenarios with gauge and bottom-tau Yukawa coupling unification.
 Indeed, for small and moderate values of $\tan \beta$
 and a physical bottom
mass $M_b = 4.9 \pm 0.3$ GeV,
the top quark mass necessary
to achieve unification of gauge and bottom-tau
Yukawa couplings is
within a ten percent of the infrared quasi fixed point
predictions for this quantity. This  result  is not modified
under the inclusion of low and high energy threshold corrections
to the gauge couplings.
In addition, we have shown that,
whenever the top quark mass is close  to its infrared quasi
fixed point value, the predictions for this quantity become
stable under variations of the bottom - tau Yukawa unification
scale as well as under small high energy threshold corrections
on the bottom and tau Yukawa couplings.
Finally, it is worth  emphasizing that the experimental
confirmation of the infrared quasi fixed point solution
will require not only the knowledge of the top quark mass, but
of $\tan\beta$ as well, an additional information which may come,
 for example, from the Higgs sector of the theory.
If confirmed experimentally,  the infrared quasi fixed point
solution will demand for understanding
the large values of the top quark Yukawa coupling
at scales $Q = {\cal{O}}(M_{GUT})$, either within the perturbative
scenario or from nonperturbative physics  at these
high energy scales.
\\
{}~\\
Acknowledgements: W.B. would like to thank members of the
Max Planck Institute for Physics for their hospitality and
support for this research.
Useful discussions with M. Olechowski are
gratefully  acknowledged. S. P. is partially supported  by the
Polish Committee for Scientific Research.
 \newpage
\parskip = 5 pt
\baselineskip = 17 pt
{}~\\
{\bf{FIGURE CAPTIONS}}\\
{}~\\
{}~\\
Fig. 1. Top quark mass  renormalization group
running, as a function of
$x = \ln(Q/M_Z)^2$, for different boundary conditions
of the top quark Yukawa coupling at
the scale $M_X = 10^{16}$ GeV,  $\alpha_3(M_Z) = 0.12$
and $\tan\beta = 5$.\\
{}~\\
Fig. 2. Bottom quark (solid line)
and Tau (dashed line) Yukawa coupling renormalization
group running as a function of the renormalization group
scale $Q$, for $\tan \beta = 3.5$ and for
a) a bottom quark mass
 $M_b = 4.9$ GeV, $\alpha_3(M_Z) = 0.11$
and  different values of the running top quark mass, which, starting
from below read
$m_t [GeV] =$  134, 154, 169, 174, 176, 177, 179, 180;
b) $M_b = 5.2$ GeV, $\alpha_3(M_Z) = 0.11$ and
$m_t [GeV] =$  134, 154, 161, 170, 173, 176, 177, 179, 180;
c) $M_b = 4.9$ GeV, $\alpha_3(M_Z) = 0.125$ and
$m_t [GeV] =$  134, 154, 174, 184, 187, 188, 189, 191, 192;
d) $M_b = 5.2$ GeV,    $\alpha_3(M_Z) = 0.125$ and
$m_t [GeV] =$  134, 154, 174, 184, 187, 188, 190.
The tau Yukawa coupling  being approximately the same for
all the above values of $m_t$.  \\
{}~\\
Fig. 3. Comparison of the infrared
quasi fixed point top quark mass predictions (solid line)
with the top quark mass necessary to achieve
bottom - tau Yukawa coupling unification
as a function of
the strong gauge coupling $\alpha_3(M_Z)$,
for $M_b = 5.2$ GeV (dot-dashed line), $M_b = 4.9$ GeV
(long dashed line) and $M_b = 4.7$ GeV (dashed line), and for
a) $\tan\beta = 1$; b) $\tan\beta = 2$; c) $\tan\beta = 5$;
d) $\tan\beta = 50$, where the dot-long dashed line represents
the values at which the three Yukawa couplings unify. Also
shown in the figure is  the band predicted from the condition
of gauge coupling unification and the experimental correlation
between $M_t$ and $\sin^2\theta_W(M_Z)$, which extends to
the right of the dash-long dashed line.
\\

\end{document}